\documentclass[11pt, lettersize]{article}

\usepackage[english]{babel}
\usepackage[utf8]{inputenc}
\usepackage[T1]{fontenc}
\usepackage{lipsum}
\usepackage{setspace}
\usepackage{booktabs}
\usepackage[document]{ragged2e}
\usepackage{csquotes}

\usepackage[a4paper,top=3cm,bottom=2cm,left=3cm,right=3cm,marginparwidth=1.75cm]{geometry}

\usepackage{amsmath}
\usepackage{graphicx}
\usepackage{tikz}
\usepackage[colorinlistoftodos]{todonotes}
\usepackage[colorlinks=true, allcolors=blue]{hyperref}
\usepackage{authblk}

\author[a]{Ho-Chun Herbert Chang$^*$}
\author[b,c]{Emily Chen$^*$}
\author[a]{Meiqing Zhang$^*$}
\author[b]{Goran Muric$^*$}
\author[a,b,c]{Emilio Ferrara}

\affil[a]{USC Annenberg School of Communication}
\affil[b]{USC Information Sciences Institute}
\affil[c]{USC Department of Computer Science}
\affil[*]{\small These authors contributed equally to this work}
\date{}
\title{Social Bots and Social Media Manipulation in 2020: \\ The Year in Review}

\begin{document}
\maketitle

\singlespacing

\section*{Abstract}

The year 2020 will be remembered for two events of global significance: the COVID-19 pandemic and 2020 U.S. Presidential Election. In this chapter, we summarize recent studies using large public Twitter data sets on these issues. We have three primary objectives. First, 
we delineate epistemological and practical considerations when combining the traditions of computational research and social science research. A sensible balance should be struck when the stakes are high between advancing social theory and concrete, timely reporting of ongoing events. We additionally comment on the computational challenges of gleaning insight from large amounts of social media data. 
Second, 
we characterize the role of social bots in social media manipulation around the discourse on the COVID-19 pandemic and 2020 U.S. Presidential Election. 
Third, we compare results from 2020 to prior years to note that, although bot accounts still contribute to the emergence of echo-chambers, there is a transition from state-sponsored campaigns to domestically emergent sources of distortion. Furthermore, issues of public health can be confounded by political orientation, especially from localized communities of actors who spread misinformation. 
We conclude that automation and social media manipulation pose issues to a healthy and democratic discourse, precisely because they distort representation of pluralism within the public sphere.

\section{Introduction}

In 2013, the World Economic Forum (WEF)'s annual \textit{Global Risk} report highlighted the multidimensional problems of misinformation in a highly connected world~\cite{WorldEconomicForum2013}. The WEF described one of the first large-scale misinformation instances that shocked America: an event from 1938, when thousands of Americans confused a radio adaptation of the H.G. Wells novel \textit{The War of the Worlds} with an official news broadcast. Many started panicking, in the belief that the United States had been invaded by Martians. 

Today, it would be hard for a radio broadcast to cause comparably widespread confusion. First, broadcasters have learned to be more cautious and responsible; and second, listeners have learned to be more savvy and sceptical. However, with social media, we are witnessing comparable phenomena on a global scale and with severe geopolitical consequences. A relatively abrupt transition from a world in which few traditional media outlets dominated popular discourse, to a  multicentric highly-connected world where information consumers and producers coalesced into one, can bring unparalleled challenges and unforeseen side effects. A sudden democratization in the media ecosystem enables everyone online to broadcast their ideas to potentially massive audiences, thus allowing content that is not necessarily moderated or curated to be broadly accessible. Extreme opinions can become increasingly more visible and fringe groups can start gaining unprecedented attention. Eccentric ideas that would otherwise garner little support within fringe communities, now could make their way into the mainstream. Furthermore, the free canvas of highly connected social media systems has been reportedly exploited by malicious actors, including foreign governments and state-sponsored groups, willing to deliberately misinform for their financial or political gain.

Nowadays, the use of social media to spread false news, provoke anxiety and incite fear for political reasons has been demonstrated around the World \cite{Bessi2016, Shao2018, badawy2019falls, Derczynski2019, Badrinathan2020, Mujani2020, Schroeder2020, luceri2019red}. However, social media manipulation is not exclusively tied to political discourse. Public health can also be endangered by the spread of false information. For instance, in January 2019, panic erupted in Mumbai schools caused by social media rumors that the vaccines were a plot by the government to sterilize Muslim children: That led to only 50\% of those who were expected to be vaccinated to actually get the vaccine~\cite{Larson2020}.

Researchers from the \textit{Democracy Fund} and \textit{Omidyar Network} in their investigative report titled ``\textit{Is Social Media a Threat to Democracy?}'',~\cite{Deb2017} warn that the fundamental principles underlying democracy ---trust, informed dialogue, a shared sense of reality, mutual consent, and participation--- are being put to the ultimate litmus test by certain features and mechanisms of social media. They point out six main issues: 1) Echo chambers, polarization, and hyper-partisanship; 2) Spread of false and/or misleading information; 3) Conversion of popularity into legitimacy; 4) Manipulation by populist leaders, governments, and fringe actors; 5) Personal data capture and targeted messaging/advertising; and 6) Disruption of the public square.

As a matter of research, these six issues can be studied through multiple academic and epistemological angles. \textit{Computational Social Science} has evolved swiftly in the past few years: Students of the social sciences are becoming masters of machine learning, while students of computer science interested in social phenomenon develop domain expertise in sociology, political science, and communication. More so than a methodological evolution, it is a shared critical interest in the growing impact social media platforms play in the very fabric of our society. 
A special issue documenting ``Dark Participation''~\cite{quandt2018dark} contrasts various issues of misinformation across different governments~\cite{quandt2021can}. Scholars point out an increasingly shared challenge: the balance of combating foreign interference without compromising domestic free speech~\cite{chang2021digital}. The resolution of these issues requires iteration between computational insights and policy-makers, as any type of intervention will inadvertently attract critiques of suppression or create unforeseen side effects.

\subsection{Focus of this Chapter}
In this chapter, we focus on spread of false and/or misleading information across two salient dimensions of social media manipulation, namely (i) automation (e.g., prevalence of bots), and (ii) distortion (misinformation, disinformation, injection of conspiracies or rumors). We provide direct insight into two case studies: a) the COVID-19 pandemic and b) the 2020 U.S. Presidential Election. We detail the many aspects of large-scale computational projects: a) tracking and cleaning billions of tweets, b) enriching the data through state-of-the-art machine learning, and c) recommendation of actionable interventions in regards to platform governance and online speech policy. 

While misleading information can materialize in many different forms, it is often scrutinized in the context of current events. Social media allows users to actively engage in discourse in real-time, reacting to breaking news and contributing to the conversation surrounding a particular topic or event with limited filters for what can or cannot be posted prior to publication. Although many social media companies have terms of services and automated filters that remove posts that violate their community guidelines, many of these posts are either able to evade detection long enough such that a wide audience has already seen or engaged with a post, or elude these automated or human-assisted filters completely. 

Politics and current events as a whole have created an environment that is rife and conducive to the spread of misleading information. Regardless of the alacrity of a post’s removal and the original poster’s broader visibility, as long as misinformation has been posted online, there is the potential for this information to have been seen and consequently consumed by others who can further disseminate it. Social media companies such as Twitter, Facebook and YouTube have recently begun active campaigns to reduce the spread of misinformation and conspiracy theories~\cite{jones2020youtube,yurieff2020twitter}. Fact checkers actively monitor rumors and events. However, the virality and speed at which this information propagates makes it difficult to catch and contain, particularly as alternative social media platforms, such as Parler and Gab, with fewer mitigation measures emerge to allow further misinformation circulation in the ecosystem~\cite{lima2018inside,aliapoulios2021early}. 

With the recent 2020 U.S. Presidential Election and ongoing COVID-19 pandemic, the need to understand the distortion of information becomes ever more urgent. When we discuss distortion of information, we note a subtle but important distinction between (a) misinformation, the \textit{organic} spread of false or inaccurate information, and (b) disinformation, the \textit{deliberate} spread of misinformation. Although the two terms are closely related, the nuance of purpose differentiates the intent of the distortion. Disinformation, in particular, is often promulgated on social media platforms not only by human users, but also by bots~\cite{shao2017spread, ferrara2018measuring, starbird2019disinformation}. A ``bot'', which is shorthand for the word ``software robot'', is a software based unit whose actions are controlled by software instead of human intervention. While there are many disciplines that leverage this term, we use the term ``bot'' in the context of ``social bots'', which are social media accounts that are either fully controlled by software or have some level of human intervention (semi-automated)~\cite{ferrara2016rise}. 

\section{Background and Framing}

The term \textit{computational social science} evokes not just two disciplines, but their own practices and traditions. 
In the following, we highlight some important epistemological concepts that inform the study of social media manipulation through the lens of computational and social science theory.



\subsection{Epistemology}
Although both inductive and deductive reasoning is common in social science research methods, quantitative social science research traditionally holds deductive methods in higher regard. A deductive approach starts from theories and uses data to test the hypotheses stemmed from the theories. Computational social science work conducted by computer scientists often exhibits a data-driven, inductive approach. However, as data science and domain expertise in the social sciences are brought together, computational social science bears great promise to reconcile inductive and deductive reasoning~\cite{evans2016machine}. Exploring large volumes of data, even sans prior theoretical assumptions, may yield new insights or surprising evidence. The findings from this initial, data-driven step will guide us to discern emerging hypotheses and collect new data to test them. This is called the \textit{abductive} analysis~\cite{timmermans2012theory}. It starts with observations, which serve to generate new hypotheses or filter existing hypotheses. The promising hypotheses emerged from data analysis can then be tested deductively with new data. 

This deductive approach can be used to study the relationship between social media and democratic discourse, which is hardly a direct or linear one. Social media do not inherently undermine or improve democracy. Instead, they affects the quality of democracy through multiple mechanisms such as political polarization and disinformation~\cite{tucker2018social}. These intermediate variables operate in varying contexts shaped by political institutions, political culture and media ecosystems. Therefore, the effects of social media on democracy differ despite the same technological affordances~\cite{benkler2018network}. The political system, ideological distribution, how political elites use social media and the behavioral patterns of different political actors in a given context interact with one another to determine whether political polarization and disinformation are amplified on social media platforms. The interactions amongst all potential political, social and technological variables form a complex system. Data exploration and analysis can help uncover crucial variables operating in a specific context. Our case studies of misinformation in the context of the COVID-19 pandemic and the 2020 U.S. Presidential Election described next will reveal significant factors underlying the relationship between social media use and democracy in the U.S. context and help identify social scientific hypotheses that are worth further investigation.

\section{Case-studies}
\subsection{Misinformation and COVID-19}

We recently found ourselves at the intersection of two important events that have changed the way the world has functioned. 2020 was already going to be a big year for U.S. politics due to the contentious nature of the current political climate. The United States has become more polarized, leading to high anticipation over whether or not the then incumbent President Trump would win re-election. While Trump cinched the Republican nomination, there was a high anticipated battle for the Democratic Presidential nominee~\cite{jacobson2020politifact}.  In the midst of the political furor, in late December 2019, the first cases of \textit{novel SARS-COV-2 Coronavirus} (whose caused disease was later named COVID-19) were reported from Wuhan, China~\cite{taylor2021timeline}. As the world began to understand the severity of the illness, whose status was later classified as a pandemic, many countries began to impose lockdowns in attempts to contain the outbreaks~\cite{wu2020coronavirus, taylor2021timeline}. 

For years, our conversations had already been shifting toward online, with the advent of social media platforms that foster environments for sharing information. Social media has also become more integrated into the fabric of political communication~\cite{romero2011differences}. With the lockdowns that closed offices and forbade gatherings, the discourse surrounding current events was pushed even further onto online platforms~\cite{hadden2020latest, hadden2020companies, fischer2020social, jungherr2016twitter}. This created a breeding ground for potential misinformation and disinformation campaigns to flourish, particularly surrounding health initiatives during a time of heightened political tensions during the 2020 U.S. Presidential Election~\cite{benkler2020mailin}. In our paper published in the \textit{Harvard Misinformation Review} special issue on \textit{U.S. Elections and Disinformation}, we study the politicization of and misinformation surrounding health narratives during this time. We found several major narratives present in our data, and further explored two health-related narratives that were highly politicized: \textit{mask wearing} and \textit{mail-in ballots}. 

\subsubsection{General Dataset}\label{sec:dataset}
We have been actively collecting and maintaining two publicly released Twitter datasets: one focusing on COVID-19 related discourse and the other on the 2020 U.S. Presidential Election \cite{Chen2020, chen2020tracking}. We began the former collection in late January 2020 and the latter in late May 2019. These tweets are collected using the Twitter streaming API, which enables us to gather tweets that match specific keywords or accounts~\cite{twitter_stream}. We note here that, at the time of this writing, the free Twitter streaming API only returns 1\% of the full Twitter data stream. Because of this limitation, we are unable to collect all tweets relevant to COVID-19 and the elections. However, the 1\% returned is still a representative sample of the discourse occurring during that day~\cite{morstatter2013sample}. 

In this particular case study, we capitalized on both our COVID-19 (v1.12) and elections (v1.3) Twitter datasets, with a focus on the time period from March 1, 2020 through August 30, 2020. At the time that this study was conducted, we had only processed our election data from March 1, 2020 onward. This timeframe covers from Super Tuesday, when a significant number of states hold their primaries, through the end of the Democratic presidential primaries. 

\subsubsection{COVID-19 and the Democratic Primaries Filtered Dataset}
We first filtered our COVID-19 dataset for keywords related to the elections, including the last names of the candidates as well as general elections-related keywords (vote, mailin, mail-in, mail in, ballot). We then conducted Latent Dirichlet allocation (LDA) to identify 8 topics present within the data, using the highest coherence score to determine the optimal number of topics~\cite{blei2003lda}. After sorting tweets into their most probable topic, we leveraged the most frequent hashtags, keywords, bigrams and trigrams to understand the narratives within each identified topic. Four broader narratives emerged: general Coronavirus discourse, lockdowns, mask wearing and mail-in balloting. We then filtered our general COVID-19 and elections dataset for tweets that contained at least one of the aforementioned elections-related keywords and a representative keyword or hashtag from the four major identified topics. This netted us a final dataset of 67,846,555 tweets, with 10,536,524 general Coronavirus tweets, 619,914 regarding lockdowns, 1,283,450  tweets on mask-wearing and 5,900,737 on mail-in balloting. 

\subsubsection{Discourse}
\begin{figure}[!htb]
\centering
\includegraphics[width =0.9\linewidth]{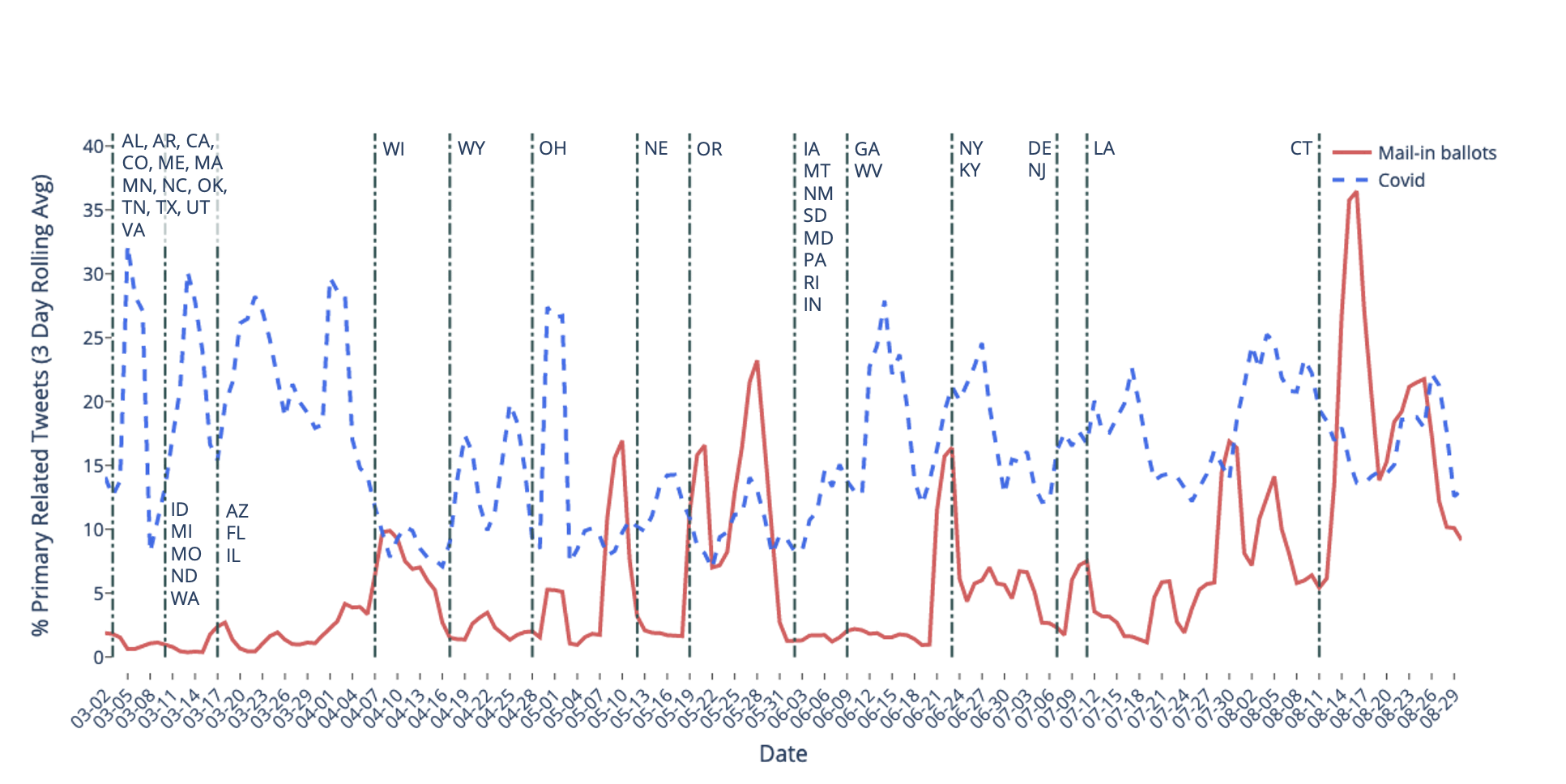}
\caption{Coronavirus and mail-in ballot related tweets within primaries related tweets, plotted as a 3-day rolling average of the percentage of primary related tweets. State abbreviations aligned with the day on which the respective state conducted their Democratic primary.} \label{fig:HKS-ballot-covid}
\end{figure}

\begin{figure}[!htb]
\centering
\includegraphics[width =0.9\linewidth]{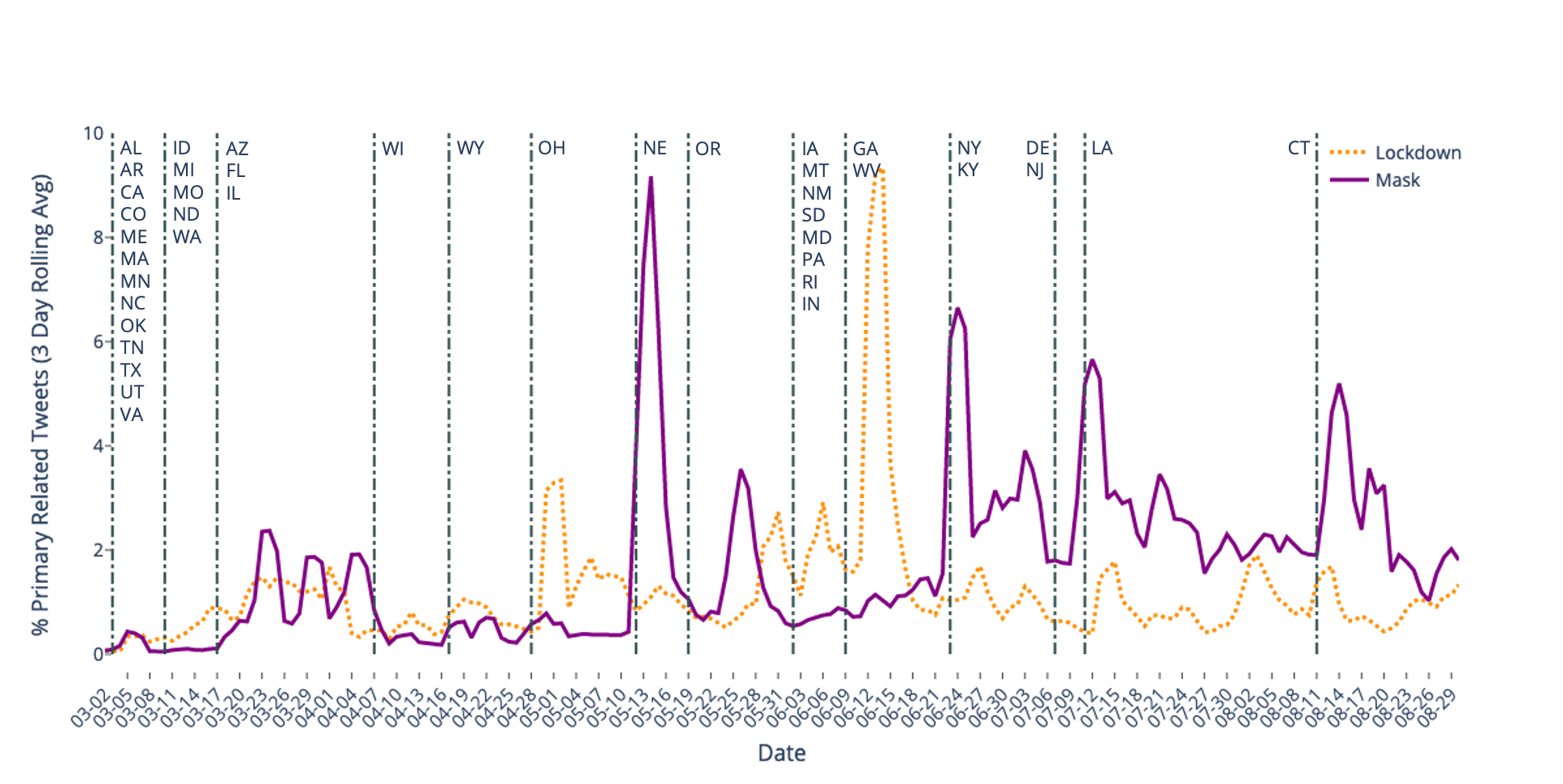}
\caption{Lockdown and mask related tweets within primaries related tweets, plotted as a 3-day rolling average of the percentage of primary related tweets. State abbreviations aligned with the day on which the respective state conducted their Democratic primary.} \label{fig:HKS-lockdown-masks}

\end{figure}

We first wanted to understand how discourse surrounding our four narratives (Coronavirus, lockdowns, mask wearing and mail in balloting) fluctuated over time (see figures \ref{fig:HKS-ballot-covid} and \ref{fig:HKS-lockdown-masks}). We tracked the percentage of all collected tweets on a particular day that contained selected keywords and hashtags that are representative of each narrative. 

\paragraph{Coronavirus.}
The pervasiveness of Coronavirus-related tweets in our Twitter dataset is by construction hence unsurprising. Not only was our COVID-19 dataset tracking Coronavirus-related keywords, but this topic has dominated political discourse in the United States since the first case was reported in Washington state on January 21, 2020. In this narrative, we find several prevalent misinformation subnarratives --- including the belief that COVID-19 is a hoax created by the Democratic party and that COVID-19 will disappear by itself~\cite{egan2020trump}. This has also been driven in tandem with the anti-vaccine movement, which has staged protests at COVID-19 vaccine distribution locations~\cite{lozano2021antivaccine}. Hydroxychloroquine (HCQ) also became a highly divisive topic within the Twitter community debating its effectiveness as treatment for COVID-19. During a press conference, then-President Trump stated that he was taking HCQ as a preventative measure~\cite{oprysko2020trump}. The \textit{United States Food and Drug Administration} (FDA) initially issued an emergency use authorization (EUA) for HCQ and the \textit{World Health Organization} included it in its treatment trials. However, the EUA was rescinded and the trials halted as results began to show that HCQ was not an effective treatment or preventative for COVID-19~\cite{edwards2020who, who2020hcq}. The controversy surrounding HCQ shows a shift in factuality surrounding the viability of HCQ, as it was initially unknown if HCQ was indeed viable. Information can develop into misinformation as its factuality changes, which further emphasizes the dangers of spreading medical information without substantive, corroborated scientific evidence. Despite evidence showing that HCQ should not be used as a treatment for COVID-19, this narrative promoting HCQ continued to spread and for many to seek this treatment. 

\paragraph{Mail-in Ballots.}
As fears surrounding COVID-19 began to grow throughout the United States, one of the major concerns with the U.S. Democratic primaries and the upcoming Presidential Election was how voters would be able to vote safely~\cite{bogage2020postmaster}. This caused many states to begin promoting mail-in ballots as a way to safely vote from home during the Democratic primaries. In August 2020, then-President Trump appointed Postmaster Louis DeJoy began reappropriating the United States Postal Service resources, making budget cuts and changing standard mail delivery protocols. This led to a significant slowdown of mail being processed and delivered, including the delivery of ballots, particularly as the U.S. began to prepare for the Presidential Election~\cite{pransky2020mail, cochrane2020postal}. 

While many were advocating for mail in ballots to be more widely used as COVID-19 precaution, others pushed the narrative that mail in ballots would increase ballot fraud. This misinformation has been proven false by fact checkers, as no evidence in previous election cycles have indicated that mail in ballots or absentee ballots increase voter fraud~\cite{farley2020trump}. This misinformation narrative that was incubating during the primaries season became an even larger misinformation campaign during the U.S. Presidential Election. 

\paragraph{Lock downs and Masking.} 
Finally, lock downs and masks were also major themes in our dataset. This is expected, as the United States began to implement social distancing ordinances, such as stay-at-home orders, in March 2020. As more states held their primaries, we see that mentions of lock downs and masks increase, suggesting that online conversation surrounding social distancing and mask wearing is driven by current events. This included misinformation narratives that claimed masks are ineffective and harmful towards one's health, when studies have shown that masks can effectively reduce COVID-19 transmission rates \cite{mueller2020quantitative, farley2020trump, khazan2020bizzare}.

\subsubsection{Echo chambers and populist leaders}
\begin{figure}[!htb]
\centering
\includegraphics[width =0.9\linewidth]{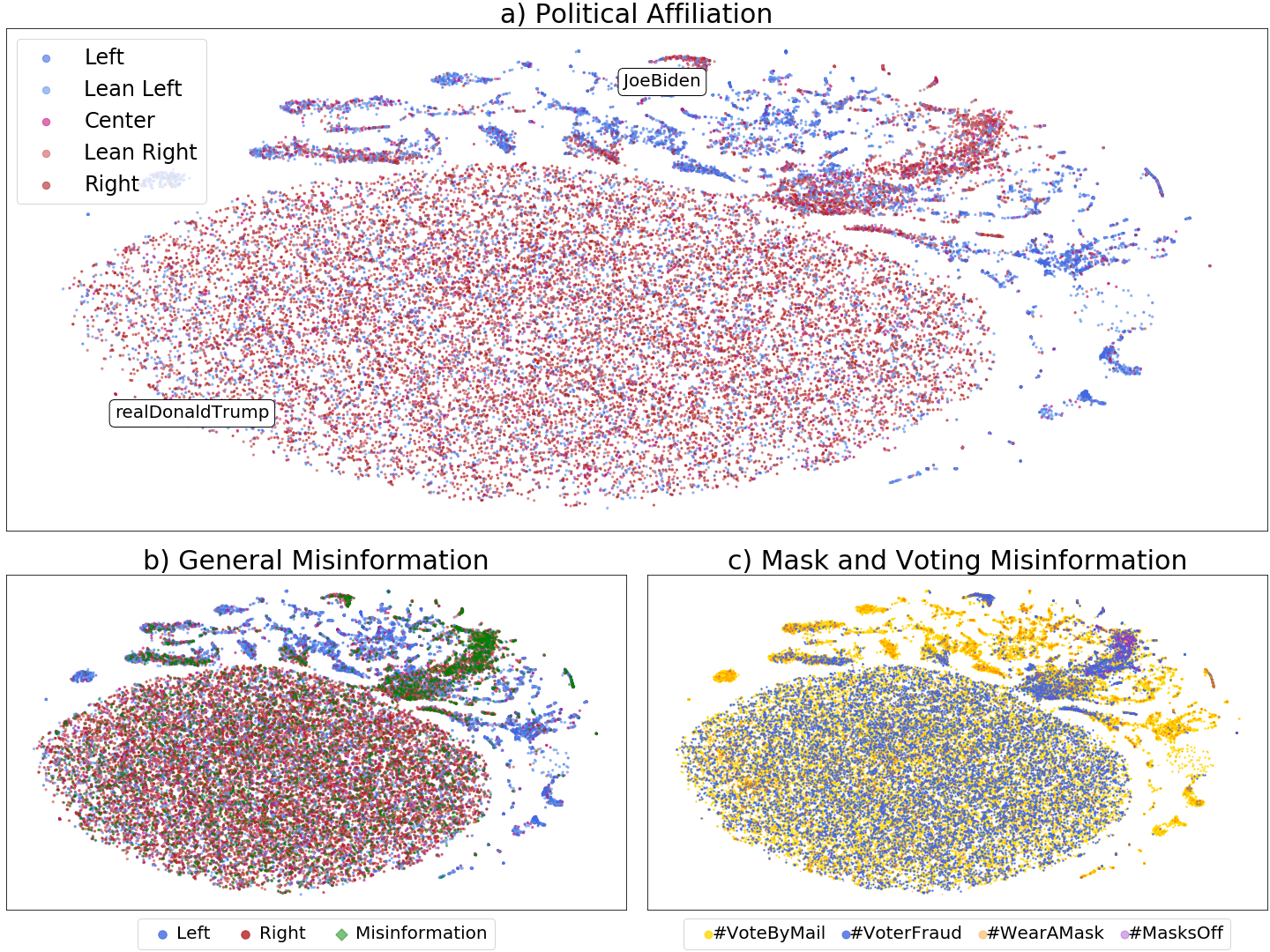}
\caption{Community structure of COVID-19 related elections discourse~\cite{Chen2021}. a) Shows the political diet of users. b) shows where general misinformation is found. c) shows the distribution of mail-in voting and mask wearing, and the position of the Twitter users. } \label{fig:HKS-Network}
\end{figure}

Out of the four narratives, we further investigate mask-wearing and mail-in balloting, as these two topics contain health-related discourse that became highly politicized and subsequently prone to misinformation. One of the more startling findings was the source of misinformation, specifically the communities in which distortions were concentrated. Figure~\ref{fig:HKS-Network} shows the network topology of Twitter users who have engaged in COVID-19 related elections discourse (see~\cite{Chen2021} for details on the methodology to generate this plot).

Figure~\ref{fig:HKS-Network}a shows the users in our dataset, each data point being colored by ``political information diet''. In order to categorize a user’s information diet, we labeled users who have shared at least 10 posts containing URLs that have been pre-tagged by the Media-Bias/Fact-Check database.\footnote{https://mediabiasfactcheck.com/} This database contains a political leanings-tagged list of commonly-shared domains (left, center-left, center, center-right and right). We found that the majority of the users are center or left-leaning. However, there is also a fairly clear distinction between more homogeneous conservative and liberal clusters near the top of the topology. This suggests that while the majority of users ingest a variety of information from both sides of the aisle, there are still clear signs of polarization based on political views that can be detected in the network topology. This polarization of highly connected clusters also indicates the presence of ``echo chambers''~\cite{jamieson2008echo, du2017echo}.  

Media-Bias/Fact-Check also contains a list of domains which they deem ``questionable sources'', or sources that are known to prompt conspiracy theories and misinformation. We use this to tag each user with both their political affiliation (Left or Right) and their tendency to spread misinformation or fact. We indicate the users who is more likely to spread misinformation in green in Figure~\ref{fig:HKS-Network}b. From this we observe that while misinformation does occur throughout the user base, conservative clusters are more likely to spread misinformation. We specifically identify a dense cluster of conservative users in the upper right hand of the topology that are more prone to engage with misinformation. 

Within the mask wearing and mail-in ballot narratives, we manually identified representative hashtags and co-occurring hashtags promoting misinformation or factual information (e.g., \#WearAMask, \#MasksOff, \#VoteByMail, \#VoterFraud). When we visualize this information on the same network topology, it is evident that there is a heterogeneity in the majority of the user’s likelihood to participate in discourse surrounding mask and mail-in ballot misinformation and fact. However, the same dense conservative cluster that we identified earlier appears to have posted tweets related to mail-in ballot and mask misinformation, compared to the left leaning clusters who tended to tweet factual information surrounding mail-in ballots and masks. Interestingly, there seems to be a divide between conservatives who push mail-in ballot misinformation and those that push mask misinformation.   

Upon closer inspection of the tweets in each cluster, we find that conservatives are not the only ones to participate in misinformation. One of the factual narratives~\cite{sherman2020politifact} that was challenged by left-leaning users was that the Obama administration had not restocked the nation’s supply of N95 masks after the H1N1 outbreak in 2009. However, the divide in misinformation narrative focus in the dense conservative cluster suggests that users within that cluster were prone to engage in misinformation about specific subjects (such as masks or mail-in ballots) instead of misinformation in general.

Our findings on the ideological patterns of misinformation on Twitter are consistent with a rising line of research that focuses on the asymmetric polarization in the U.S. context: Some political scientists argue that party polarization in the U.S. is asymmetrical, with Republicans moving more to the right than Democrats to the left~\cite{carmines2011class,hacker2006off,theriault2013gingrich}. This trend was evolving even before the advent of social media. The existing ideological asymmetry affects the exposure to media sources on digital platforms~\cite{faris2017partisanship,brady2019ideological} and leads to asymmetrical consumption of misinformation~\cite{tucker2018social}. It lends support to the existing asymmetric polarization hypothesis and highlights its important role in mediating the relationship between social media and democracy in the United States. 

\subsection{Misinformation and 2020 U.S. Presidential Election}
There is a well known saying that ``the first casualty of war is truth''. 
In times of unusual social tensions caused by the political struggle with relatively high stakes, the proliferation of false news, misinformation and other sorts of media manipulation is to be expected. The importance of voter competence is one of the postulates of modern democracy~\cite{gant,Stucki2018} and information vacuums can undermine electoral accountability~\cite{ashworth}. An ideal democracy assumes an informed and rational voter, but the former aspect is something that can be undermined or compromised. During the 2020 U.S. Presidential Election, social media manipulation has been observed in the form of (i) automation, that is the evidence for adoption of automated accounts governed predominantly by software rather than human users, and (ii) distortion, in particular of salient narratives of discussion of political events, e.g., with the injection of inaccurate information, conspiracies or rumors.
In the following, we describe ours and others' findings in this context.

\subsubsection{Dataset}
For this study, we again leverage one of our ongoing and publicly released Twitter datasets centered around the 2020 U.S. Presidential Election. Please refer to Section~\ref{sec:dataset} for more details on the collection methods; this particular dataset is further described in ~\cite{Chen2020}. While this dataset now has over 1.2 billion tweets, we focused on tweets posted between June 20, 2020 and September 9, 2020 in advance of the November 3, 2020 election. This subset yielded 240 million tweets and 2 TB of raw data. The period of observation includes several salient real-world political events, such as the \textit{Democratic National Committee} (DNC) and \textit{Republican National Committee} (RNC) conventions. 

\subsubsection{Automation Detection}
The term \textit{bot} (shorthand for robot) in Computational Social Science commonly refers to fully automated or semi-automated accounts on social media platforms~\cite{ferrara2016rise}. Research into automation on social media platforms has spurned its own sub-field not only in computational social sciences but in social media research at large~\cite{ferrara2016rise, shao2017spread, aiello2012people, zelenkauskaite2017information, yang2019arming}. One of the major challenges with automation is the ability to detect accounts that are bots as opposed to accounts fully operated by humans. Although there are benign accounts that publicly advertise the fact that they are automated, bots used for malicious purposes try to evade detection. As platforms and researchers study the behavior of bots and devise algorithms and systems that are able to automatically flag accounts as bots, bot developers are also actively developing new systems to subvert these detection attempts by mimicking behavioral signals of human accounts~\cite{sayyadiharikandeh2020detection, yang2020scalable}

Botometer is a tool developed and released by researchers at \textit{Indiana University}, as part of the \textit{Observatory on Social Media} (OSoMe~\cite{davis2016osome}), that allows users to input a Twitter user's screen name, and returns a score of how likely an account is to be automated.\footnote{https://botometer.osome.iu.edu/} These scores range from 0 to 5, with 0 indicating that the account has been labeled as most likely human and 5 indicating that the account is most likely a bot account. We will be referring to accounts that are most likely human accounts as ``human'' and bot-like accounts as ``bots'' for brevity. Botometer itself has gone through several iterations, with the most recent version Botometer v4 released in September 2020~\cite{sayyadiharikandeh2020detection}. Botometer v4 extracts thousands of features from an input account and leverages machine learning models trained on a large repository of labeled tweets to predict the likelihood of an account being a bot. Botometer v4 \cite{yang2020scalable} can identify different types of bots, including bots that are fake followers, spammers and astroturfers~\cite{yang2019arming, ferrara2019history}.

\subsubsection{Automation in social media manipulation during 2020 U.S. Presidential Election}

In the following analysis, we leveraged Botometer v3~\cite{yang2019arming}, as that was the latest version at the time we performed our study~\cite{Ferrara2020}. We tagged 32 percent of the users within our complete dataset, and removed all tweets not posted by the users for whom we have bot scores for. We labeled the top decile of users according to Botometer scores as ``bots'' and the bottom decile as ``humans''~\cite{Ferrara2020_1}. Our final dataset contains more that four million tweets posted by bots and more than one million tweets posted by humans. We found that a number of the top hashtags used in tweets by bots are affiliated with well known conspiracy theories that will be studied later in this chapter (e.g., \#wwg1wga, \#obamagate, \#qanon) and others are Trump’s campaign related hashtags. In contrast, tweets from humans contain a mix of both Trump and Biden campaign hashtags. 

We use campaign-related hashtags in order to distinguish between users who engage in left-leaning (Biden campaign) and right-leaning (Trump campaign) political discourse. 
We find that there are over 2.5 million left-leaning humans, and a little over 18,000 left-leaning bots. Comparatively, we found over 8.5 million right-leaning humans and almost 85,000 right-leaning bots. This enables us to take a snapshot of how right-leaning bots and humans engage in election-related narratives compared to their left-leaning counterparts. What is interesting here is whether or not there are distinguishable features of bots and humans based on their political affiliations and engagements within the network~\cite{luceri2019red}. 

What we find is that right leaning bots tend to post right-leaning news, with many accounts also posting highly structured (i.e., templated, or copy-pasted) tweets. When we manually inspected a random sample of these tweets, we found that these tweets contained similar combinations of hashtags and oftentimes similarly structured content. Many of the tweets also contained URLs to well known conspiracy news websites. Right-leaning bots also tended to have higher bot scores compared to their left-leaning counterparts, suggesting a more profound use of automation. A manual inspection of a random set of left-leaning bot tweets found that these tweets are significantly less structured, exhibiting fewer automation cues. Although disambiguation by means of specific campaign-related hashtags is not perfect, prior studies investigating political polarization has shown that the vast majority of users posting campaign-specific hashtags align with the same political party~\cite{jiang2020political,bail2018exposure}. We also find that the bot scores for bots range from 0.485 through 0.988, suggesting that the broad range of scores captures accounts that are hybrid accounts, partially automated and partially controlled by humans. 

\begin{figure}[!htb]
\centering
\includegraphics[width =0.9\linewidth]{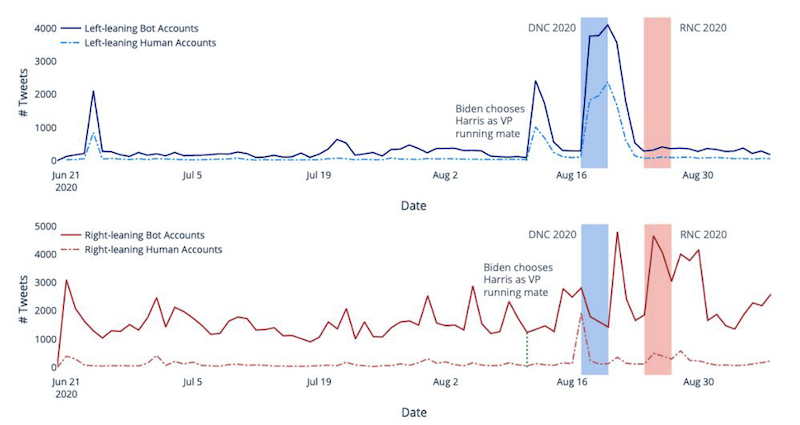}
\caption{Time series of activity of bot vs human accounts with political affiliation~\cite{Ferrara2020}.} \label{fig:FM-TS}
\end{figure}

When isolating the activity of these bot and human accounts and then examining their temporal activity, we see that each group behaves differently. Despite being outnumbered by several orders of magnitude, just a few thousand bots generated spikes of conversations around real-world political events comparable with the volume of activity of humans~\cite{Ferrara2020}. We find that conservatives, both bot and humans, tend to tweet more regularly than liberal users. The more interesting question, beyond raw volume, is whether bots play a community role in polarization. 

We found both surprising similarities and stark differences across the partisan divide. Figure~\ref{fig:FM-TS} shows the discourse volume of the top 10\% of bots and top 10\% of humans, split between left-leaning accounts (top) and right-leaning accounts (bottom). Although bots tweet in higher volumes in both cases, the activities of left-leaning bots are more localized to specific events. In contrast, right-leaning bots generate large amounts of discourse in general, showing high level of background activity.

\begin{figure}[!htb]
\centering
\includegraphics[width =0.9\linewidth]{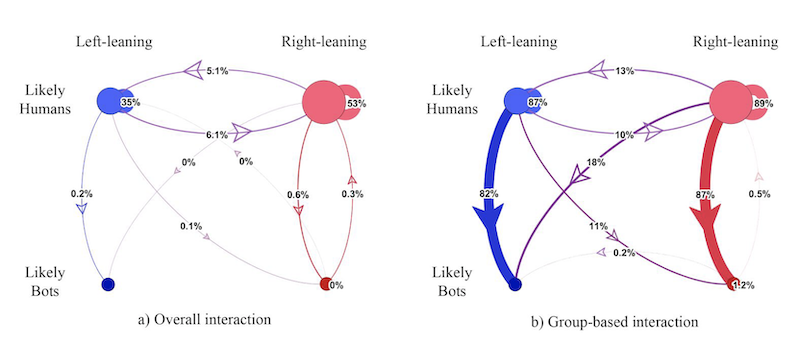}
\caption{Meso-flow of bot and human accounts by political leaning. a) Total volume of retweets between the four groups. b) Relative volume of retweets between the groups.} \label{fig:FM-meso}
\end{figure}

Next, we illustrate how do these four groups interact with each other. Figure~\ref{fig:FM-meso} shows the interactions between human and bot accounts divided by political leaning. Bots predominantly retweet humans from within their own party lines, whereas humans retweet other humans from within their party lines. At a relative retweet rate within the same party as more than 80\%, this indicates a significant level of political polarization.


\subsubsection{Distortion in social media manipulation during 2020 U.S. Presidential Election}
Next, we broaden an analysis to \textit{distortion}, an umbrella concept that also includes completely fabricated narratives that do not have a hold in reality. 
Fake news are an example of distorted narratives and are conceptualized as \textit{distorted signals uncorrelated with the truth}~\cite{Allcott2017}. To avoid the conundrum of establishing what is true and what is false to qualify a piece of information as fake news (or not), in this study we focus on conspiracy theories, another typical example of distorted narratives. Conspiracy theories can be (and most often are) based upon falsity, rumors, or unverifiable information that resist falsification; other times they are instead postulated upon rhetoric, divisive ideology, and circular reasoning based on prejudice or uncorroborated (but not necessarily false) evidence. Conspiracies can be shared by users or groups with the aim to deliberately deceive or indoctrinate unsuspecting individuals who genuinely believe in such claims~\cite{VanProoijen2019}. 

Conspiracy theories are attempts to explain the ultimate causes of significant social and political events and circumstances with claims of secret plots by powerful actors. While often thought of as addressing governments, conspiracy theories could accuse any group perceived as powerful and malevolent~\cite{Douglas2019}. They evolve and change over time, depending on the current important events. Upon manual inspection, we found that some of the most prominent conspiracy theories and groups in our dataset revolve around topics such as: objections to vaccinations, false claims related to 5G technology, a plethora of Coronavirus related false claims and the flat earth movement~\cite{Ferrara2020_1}. Opinion polls carried out around the world reveal that substantial proportions of population readily admit to believing in some kind of conspiracy theories~\cite{Byford2011}. In the context of democratic processes including the 2020 U.S. Presidential Election, the proliferation of political conspiratorial narratives could have an adverse effect on political discourse and democracy.

In our analysis, we focused on three main conspiracy groups:

\begin{enumerate}
    \item \textbf{QAnon conspiracies:} A far-right conspiracy movement whose theory suggests that President Trump has been battling against a Satan worshipping global child sex-trafficking ring and an anonymous source called 'Q' is cryptically providing secret information about the ring~\cite{Zuckerman2019}. The users who support such ideas frequently use hashtags such as \#qanon, \#wwg1wga (where we go one, we go all), \#taketheoath, \#thegreatawakening and \#qarmy. The examples of a typical tweet from the QAnon supporters are: 
    \begin{displayquote}
    \textit{''@potus @realDonaldTrump was indeed correct,the beruit fire was hit by a missile, oh and to the rest of you calling this fake,you are not a qanon you need to go ahead and change to your real handles u liberal scumbags just purpously put out misinfo and exposed yourselves,thnxnan''}
    \end{displayquote}
    
    \begin{displayquote}
    \textit{''I've seen enough. It's time to \#TakeTheOath There's no turning back now. We can and only will do this together. \#WWG1WGA \#POTUS @realDonaldTrump \#Qanon''}
    \end{displayquote}
    
    \item \textbf{``gate'' conspiracies:} Another indicator of conspiratorial content is signalled by the suffix '-gate' with theories such as pizzagate, a debunked claim that connects several high-ranking Democratic Party officials and U.S. restaurants with an alleged human trafficking and child sex ring. The examples of the typical conspiratorial tweets related to these two conspiracies are: 
    \begin{displayquote}
    \textit{''\#obamagate when will law enforcement take anything seriously? there is EVIDENCE!!!! everyone involved in the trafficking ring is laughing because they KNOW nothing will be done. @HillaryClinton @realDonaldTrump. justice will be served one way or another. literally disgusting.''}
    \end{displayquote}
    
    \begin{displayquote}
    \textit{''\#Obama \#JoeBiden, \& their top intel officers huddled in the Oval Office shortly before @realDonaldTrump was inaugurated to discuss what they would do about this new president they despised, @TomFitton in Breitbart. Read:...''}
    \end{displayquote}
    
    \item \textbf{Covid conspiracies:} A plethora of false claims related to the Coronavirus emerged right after the pandemic was announced. They are mostly related to scale of the pandemic and the origin, prevention, diagnosis, and treatment of the disease. The false claims typically go alongside the hashtags such as \#plandemic, \#scandemic or \#fakevirus. The typical tweets referring to the false claims regarding the origins of the Coronavirus are: 
    \begin{displayquote}
    \textit{''@fyjackson @rickyb\_sports @rhus00 @KamalaHarris @realDonaldTrump The plandemic is a leftist design. And it's backfiring on them. We've had an effective treatment for COVID-19, the entire time. Leftists hate Trump so much, they are willing to murder 10's of thousands of Americans to try to make him look bad. The jig is up.''}
    \end{displayquote}
    
    \begin{displayquote}
    \textit{''The AUS Govt is complicit in the global scare \#Plandemic. They are scarifying jobs, businesses freedom and families in an attempt to stop @realDonaldTrump from being reelected. Why?''}
    \end{displayquote}
    
\end{enumerate}

During the period preceding the 2020 U.S. Presidential Election, \textit{QAnon} related material has more highly active and engaged users than other narratives. This is measured by the average number of tweets an active user has made on a topic. For example, the most frequently used hashtag, \#wwg1wga, had more than 600K tweets from 140K unique users; by contrast \#obamagate had 414K tweets from 125K users. This suggests that the \textit{QAnon} community has a more active user base strongly dedicated to the narrative. 

When we analyze how the conspiratorial narratives are endorsed by the users, conditioned upon where they fall on the political spectrum, we discover that conspiratorial ideas are strongly skewed to the right. Almost a quarter of users who endorse predominantly right-leaning media platforms are likely to engage in sharing conspiracy narratives. Conversely, out of all users who endorse left-leaning media, approximately two percent are likely to share conspiracy narratives. 

Additionally, we explore the usage of conspiracy language among automated accounts. Bots can appear across the political spectrum and are likely to endorse polarizing views. Therefore, they are likely to be engaged in sharing heavily discussed topics including conspiratorial narratives. Around 13\% of Twitter accounts that endorse some conspiracy theory are likely bots. This is significantly more than users who never share conspiracy narratives, which have only 5\% of automated accounts. It is possible that such observations are in part the byproduct of the fact that bots are programmed to interact with more engaging content, and inflammatory topics such as conspiracy theories provide fertile ground for engagement~\cite{Stella2018}. On the other hand, bot activity can inflate certain narratives and make them popular. 


The narratives of these conspiracy theories during the 2020 U.S. Presidential Election call attention to the so-called ``new conspiracism'' and the partisan differences in practicing it~\cite{rosenblum2020lot}. Rosenblum and Muirhead argue that the new conspiracism in the contemporary age is ``conspiracy without theory''. Whereas the ``classic conspiracy theory'' still strives to collect evidence, find patterns and logical explanations to construct a ``theory'' of how malignant forces are plotting to do harm, the new conspiracism skips the burdens of ``theory construction'' and advances itself by bare assertion and repetition~\cite{rosenblum2020lot}. Repetition produces familiarity, which in turn increases acceptance~\cite{paul2016russian,lewandowsky2012misinformation}. A conspiracy becomes credible to its audience, simply because many people are repeating it~\cite{rosenblum2020lot}. The partisan asymmetry in the circulation of conspiracy theories is also consistent with others’ claims that the new conspiracism is asymmetrically aligned with the radical right in the U.S. context~\cite{benkler2018network,rosenblum2020lot}, although this species of conspiracism is not ideologically attached to liberals or conservatives~\cite{rosenblum2020lot}. Our analysis shows the promising direction of testing the theories of asymmetrical polarization and exploring the nature and consequences of asymmetrical media ecosystem, ideally using multi-platform data.  

The findings about the bot behaviors relative to humans on Twitter reveal some patterns of conspiracy transmission in the 2020 U.S. Presidential Election. Their high-volume and echo-chamber retweeting activities attest to the role that automation plays in stoking the new conspiracism. Bots are capable of retweeting and repeating the same information efficiently. However, bots are not solely to blame for the prevalence of conspiracy-theory stories. False information are found to spread faster than true information due to the human tendency to retweet it. A comprehensive study conducted by Vosoughi \textit{et al.} compared the diffusion of verified true and false news stories on Twitter from 2006 to 2017. They discovered that falsity travels wider and deeper than truth, even after bots were removed, suggesting that humans are more likely to retweet false rumors than true information. Among all topics, political rumors are particularly viral. False rumors peaked before and around the 2012 and 2016 U.S. Presidential Election~\cite{vosoughi2018spread}. Additionally, automated accounts that are part of an organized campaign can purposely propel some of the conspiracy narratives, further polarizing the political discourse.

Although bots present a threat to the ideal, well-informed democratic citizenship, the susceptibility of humans to believing and spreading false information is worth equal attention. Further examinations of how distorted narratives go viral will help us better diagnose the problem. Some new research points to the hypothesis that the nature and structure of false rumors and conspiracy-theory stories evoke human interest. For example, Vosoughi \textit{et al.} suggested that false rumors tend to be more novel, hence more salient.
False rumors also elicit stronger emotions of surprise and disgust~\cite{vosoughi2018spread}. Tangherlini \textit{et al.} studied the conspiracy theory narrative framework using the cases of \textit{Bridgegate} and \textit{Pizzagate}. They deconstructed those stories into multi-scale narrative networks and found that conspiracy theories are composed of a small number of entities, multiple interconnected domains and separable disjoint subgraphs. By construction, conspiracy theories can form and stabilize faster. In contrast, the unfolding of true conspiracy stories will admit new evidence and result in a denser network over time~\cite{tangherlini2020automated}. Therefore, true stories could be at a disadvantage when competing with false rumors as they are less stable and grow in complexity as events develop.

\section{Conclusions}
In this chapter, we presented the findings that emerged from two significant events of 2020. In the first study, we showed how political identity aligns with narratives of public health. Four narratives were identified: (i) mail-in ballots, (ii) reference to the pandemic, (iii) lock-downs, and (iv) mask-wearing. Spikes in these narratives were found to be driven by predetermined events, predominantly the primaries. When observing the policy stance of mail-in ballots and mask-wearing, we observe users against mask-wearing and mail-in ballots arise from a dense group of conservative users separate from the majority. Topological distinctions between these two groups are further observed. Further details are found in our recent paper~\cite{chang2021digital}.

When investigating the 2020 U.S. Presidential Election more broadly, we find bots not only generate much higher volumes of election-related tweets per capita, but also tweet primarily within their own political lines (more than 80\% for both left- and right-leaning communities). An analysis of content from QAnon-driven conspiracies, politicized ``gate''-related, and COVID-related conspiracies suggested that users self-organize to promulgate false information and also leverage automation to amplify hyperpartizan and conspiratorial news sites: more details are discussed in our associated study~\cite{Ferrara2020_1}.

What do these results tell us? First, although bots still generate significant distortions in volume and self-reinforcement across party lines as observed in the 2016 U.S. Presidential Election~\cite{Bessi2016}, this is overshadowed by the self-organization of extremism and ``new conspiracism'' in the public sphere. A further contrast is the shift from foreign interference in 2016 to domestic, ingrown social media manipulation in 2020. This phenomenon can be observed across a variety of case studies, including the populism in EU~\cite{muis2017causes}, xenophobia in Russia, hate speech in Germany~\cite{zhuravskaya2020political}, and foreign interference in Taiwan~\cite{chang2021digital}.

Finally, the case study of COVID-19 demonstrates the interplay between public health and politics on a national level. In the past, computational studies on anti-vaccination focused on smaller, community level scales~\cite{lozano2021antivaccine}. Given the high levels of alignment between political information diet and health misinformation, the polarization and subsequent distortions not only can have ramifications on the democratic process, but also tangible effects on public health.



\bibliographystyle{unsrt}
\bibliography{sample}

\begin{thebibliography}{10}

\bibitem{WorldEconomicForum2013}
{World Economic Forum}.
\newblock {Global Risks 2013}.
\newblock Technical report, World Economic Forum, 2013.

\bibitem{Bessi2016}
Alessandro Bessi and Emilio Ferrara.
\newblock {Social bots distort the 2016 U.S. Presidential election online
  discussion}.
\newblock {\em First Monday}, 21(11), nov 2016.

\bibitem{Shao2018}
Chengcheng Shao, Pik-Mai Hui, Lei Wang, Xinwen Jiang, Alessandro Flammini,
  Filippo Menczer, and Giovanni~Luca Ciampaglia.
\newblock {Anatomy of an online misinformation network}.
\newblock {\em PLOS ONE}, 13(4):e0196087, apr 2018.

\bibitem{badawy2019falls}
Adam Badawy, Kristina Lerman, and Emilio Ferrara.
\newblock Who falls for online political manipulation?
\newblock In {\em Companion Proceedings of The 2019 World Wide Web Conference},
  pages 162--168, 2019.

\bibitem{Derczynski2019}
Leon Derczynski, Torben~Oskar Albert-Lindqvist, Marius~Ven{\o} Bendsen, Nanna
  Inie, Viktor~Due Pedersen, and Jens~Egholm Pedersen.
\newblock {Misinformation on Twitter During the Danish National Election: A
  Case Study}.
\newblock TTO Conference Ltd, oct 2019.

\bibitem{Badrinathan2020}
Sumitra Badrinathan.
\newblock {Educative Interventions to Combat Misinformation: Evidence From a
  Field Experiment in India}.
\newblock aug 2020.

\bibitem{Mujani2020}
{Who Believed Misinformation during the 2019 Indonesian Election?}
\newblock {\em Asian Survey}, 60(6):1029--1043, dec 2020.

\bibitem{Schroeder2020}
Ralph Schroeder.
\newblock {Even in Sweden?}
\newblock {\em Nordic Journal of Media Studies}, 2(1):97--108, jun 2020.

\bibitem{luceri2019red}
Luca Luceri, Ashok Deb, Adam Badawy, and Emilio Ferrara.
\newblock Red bots do it better: Comparative analysis of social bot partisan
  behavior.
\newblock In {\em Companion Proceedings of the 2019 World Wide Web Conference},
  pages 1007--1012, 2019.

\bibitem{Larson2020}
Heidi~J. Larson.
\newblock {\em {Stuck - How Vaccine Rumors Start and Why They Don't go Away}}.
\newblock Oxford University Press, 1 edition, 2020.

\bibitem{Deb2017}
Anamitra Deb, Stacy Donohue, and Tom Glaisyer.
\newblock {Is Social Media a Threat to Democracy?}
\newblock Technical report, Omydiar Group, 2017.

\bibitem{quandt2018dark}
Thorsten Quandt.
\newblock Dark participation.
\newblock {\em Media and Communication}, 6(4):36--48, 2018.

\bibitem{quandt2021can}
Thorsten Quandt.
\newblock Can we hide in shadows when the times are dark?
\newblock {\em Media and Communication}, 9(1):84--87, 2021.

\bibitem{chang2021digital}
Ho-Chun~Herbert Chang, Samar Haider, and Emilio Ferrara.
\newblock Digital civic participation and misinformation during the 2020
  taiwanese presidential election.
\newblock {\em Media and Communication}, 9(1):144--157, 2021.

\bibitem{jones2020youtube}
Roy Cellan-Jones.
\newblock Youtube, facebook and twitter align to fight covid vaccine
  conspiracies, Nov 2020.

\bibitem{yurieff2020twitter}
Kaya Yurieff.
\newblock How twitter, facebook and youtube are handling election
  misinformation, Nov 2020.

\bibitem{lima2018inside}
L.~{Lima}, J.~C.~S. {Reis}, P.~{Melo}, F.~{Murai}, L.~{Araujo}, P.~{Vikatos},
  and F.~{Benevenuto}.
\newblock Inside the right-leaning echo chambers: Characterizing gab, an
  unmoderated social system.
\newblock In {\em 2018 IEEE/ACM International Conference on Advances in Social
  Networks Analysis and Mining (ASONAM)}, pages 515--522, 2018.

\bibitem{aliapoulios2021early}
Max Aliapoulios, Emmi Bevensee, Jeremy Blackburn, Emiliano~De Cristofaro,
  Gianluca Stringhini, and Savvas Zannettou.
\newblock An early look at the parler online social network, 2021.

\bibitem{shao2017spread}
Chengcheng Shao, Giovanni~Luca Ciampaglia, Onur Varol, Alessandro Flammini, and
  Filippo Menczer.
\newblock The spread of fake news by social bots.
\newblock {\em arXiv preprint arXiv:1707.07592}, 96:104, 2017.

\bibitem{ferrara2018measuring}
Emilio Ferrara.
\newblock Measuring social spam and the effect of bots on information diffusion
  in social media.
\newblock In {\em Complex spreading phenomena in social systems}, pages
  229--255. Springer, 2018.

\bibitem{starbird2019disinformation}
Kate Starbird.
\newblock Disinformation's spread: bots, trolls and all of us.
\newblock {\em Nature}, 571(7766):449--450, 2019.

\bibitem{ferrara2016rise}
Emilio Ferrara, Onur Varol, Clayton Davis, Filippo Menczer, and Alessandro
  Flammini.
\newblock The rise of social bots.
\newblock {\em Commun. ACM}, 59(7):96–104, June 2016.

\bibitem{evans2016machine}
James~A Evans and Pedro Aceves.
\newblock Machine translation: Mining text for social theory.
\newblock {\em Annual Review of Sociology}, 42:21--50, 2016.

\bibitem{timmermans2012theory}
Stefan Timmermans and Iddo Tavory.
\newblock Theory construction in qualitative research: From grounded theory to
  abductive analysis.
\newblock {\em Sociological theory}, 30(3):167--186, 2012.

\bibitem{tucker2018social}
Joshua~A Tucker, Andrew Guess, Pablo Barber{\'a}, Cristian Vaccari, Alexandra
  Siegel, Sergey Sanovich, Denis Stukal, and Brendan Nyhan.
\newblock Social media, political polarization, and political disinformation: A
  review of the scientific literature.
\newblock {\em Political polarization, and political disinformation: a review
  of the scientific literature (March 19, 2018)}, 2018.

\bibitem{benkler2018network}
Yochai Benkler, Robert Faris, and Hal Roberts.
\newblock {\em Network propaganda: Manipulation, disinformation, and
  radicalization in American politics}.
\newblock Oxford University Press, 2018.

\bibitem{jacobson2020politifact}
Louis Jacobson.
\newblock Politifact - the record-setting 2020 democratic primary field: What
  you need to know, Jan 2020.

\bibitem{taylor2021timeline}
Derrick~Bryson Taylor.
\newblock A timeline of the coronavirus pandemic, Jan 2021.

\bibitem{wu2020coronavirus}
Jiachuan Wu, Savannah Smith, Mansee Khurana, Corky Siemaszko, and Brianna
  DeJesus-Banos.
\newblock Coronavirus lockdowns and stay-at-home orders across the u.s., Apr
  2020.

\bibitem{romero2011differences}
Daniel~M. Romero, Brendan Meeder, and Jon Kleinberg.
\newblock Differences in the mechanics of information diffusion across topics:
  idioms, political hashtags, and complex contagion on twitter.
\newblock {\em Proceedings of the 20th international conference on World wide
  web - WWW '11}, 2011.

\bibitem{hadden2020latest}
Joey Hadden and Laura Casado.
\newblock Here are the latest major events that have been canceled or postponed
  because of the coronavirus outbreak, including the 2020 tokyo olympics,
  burning man, and the 74th annual tony awards, Apr 2020.

\bibitem{hadden2020companies}
Joey Hadden, Laura Casado, Tyler Sonnemaker, and Taylor Borden.
\newblock 21 major companies that have announced employees can work remotely
  long-term, Dec 2020.

\bibitem{fischer2020social}
Sara Fischer.
\newblock Social media use spikes during pandemic, Apr 2020.

\bibitem{jungherr2016twitter}
Andreas Jungherr.
\newblock Twitter use in election campaigns: A systematic literature review.
\newblock {\em Journal of Information Technology \& Politics}, 13(1):72--91,
  2016.

\bibitem{benkler2020mailin}
Yochai Benkler, Casey Tilton, Bruce Etling, Hal Roberts, Justin Clark, Robert
  Faris, Jonas Kaiser, and Carolyn Schmitt.
\newblock Mail-in voter fraud: Anatomy of a disinformation campaign.
\newblock {\em SSRN Electronic Journal}, 2020.

\bibitem{Chen2020}
Emily Chen, Ashok Deb, and Emilio Ferrara.
\newblock {{\#}Election2020: The First Public Twitter Dataset on the 2020 US
  Presidential Election}.
\newblock {\em arXiv}, oct 2020.

\bibitem{chen2020tracking}
Emily Chen, Kristina Lerman, and Emilio Ferrara.
\newblock Tracking social media discourse about the covid-19 pandemic:
  Development of a public coronavirus twitter data set.
\newblock {\em JMIR Public Health and Surveillance}, 6(2):e19273, 2020.

\bibitem{twitter_stream}
Consuming streaming data | twitter developer.

\bibitem{morstatter2013sample}
Is the sample good enough? comparing data from twitter’s streaming api with
  twitter’s firehose.
\newblock 7.

\bibitem{blei2003lda}
David~M. Blei, Andrew~Y. Ng, and Michael~I. Jordan.
\newblock Latent dirichlet allocation.
\newblock {\em J. Mach. Learn. Res.}, 3(null):993–1022, March 2003.

\bibitem{egan2020trump}
Lauren Egan.
\newblock Trump calls coronavirus democrats' 'new hoax', Feb 2020.

\bibitem{lozano2021antivaccine}
Alicia~Victoria Lozano.
\newblock Anti-vaccine protest briefly shuts down dodger stadium vaccination
  site, Jan 2021.

\bibitem{oprysko2020trump}
Caitlin Oprysko.
\newblock Trump says he's taking hydroxychloroquine, despite scientists'
  concerns, May 2020.

\bibitem{edwards2020who}
Erika Edwards.
\newblock World health organization halts hydroxychloroquine study, Jun 2020.

\bibitem{who2020hcq}
Who discontinues hydroxychloroquine and lopinavir/ritonavir treatment arms for
  covid-19, Jul 2020.

\bibitem{bogage2020postmaster}
Jacob Bogage, Lisa Rein, and Josh Dawsey.
\newblock Postmaster general eyes aggressive changes at postal service after
  election, Aug 2020.

\bibitem{pransky2020mail}
Noah Pransky.
\newblock U.s. mail slowed down just before the election. these states are most
  at risk, Oct 2020.

\bibitem{cochrane2020postal}
Emily Cochrane, Hailey Fuchs, Kenneth~P Vogel, and Jessica Silver-Greenberg.
\newblock Postal service suspends changes after outcry over delivery slowdown,
  Aug 2020.

\bibitem{farley2020trump}
Robert Farley.
\newblock Trump's latest voter fraud misinformation, Apr 2020.

\bibitem{mueller2020quantitative}
Amy~V. Mueller, Matthew~J. Eden, Jessica~M. Oakes, Chiara Bellini, and
  Loretta~A. Fernandez.
\newblock Quantitative method for comparative assessment of particle removal
  efficiency of fabric masks as alternatives to standard surgical masks for
  ppe.
\newblock {\em Matter}, 3(3):950--962, 2020.

\bibitem{khazan2020bizzare}
Olga Khazan.
\newblock How a bizarre claim about masks has lived on for months, Oct 2020.

\bibitem{Chen2021}
Emily Chen, Herbert Chang, Ashwin Rao, Emilio Ferrara, Kristina Lerman, and
  Geoffrey Cowan.
\newblock {COVID-19 misinformation and the 2020 U.S. presidential election}.
\newblock {\em Harvard Kennedy School Misinformation Review}, Feb 2021.

\bibitem{jamieson2008echo}
Kathleen~Hall Jamieson and Joseph~N. Cappella.
\newblock {\em Echo chamber: Rush Limbaugh and the conservative media
  establishment}.
\newblock Oxford University Press, 2008.

\bibitem{du2017echo}
Siying Du and Steve Gregory.
\newblock The echo chamber effect in twitter: does community polarization
  increase?
\newblock In Hocine Cherifi, Sabrina Gaito, Walter Quattrociocchi, and
  Alessandra Sala, editors, {\em Complex Networks {\&} Their Applications V},
  pages 373--378, Cham, 2017. Springer International Publishing.

\bibitem{sherman2020politifact}
Amy Sherman.
\newblock Politifact - trump said the obama admin left him a bare stockpile.
  wrong, Apr 2020.

\bibitem{carmines2011class}
Edward~G Carmines.
\newblock Class politics, american-style: A discussion of winner-take-all
  politics: How washington made the rich richer—and turned its back on the
  middle class.
\newblock {\em Perspectives on Politics}, 9(3):645--647, 2011.

\bibitem{hacker2006off}
Jacob~S Hacker and Paul Pierson.
\newblock {\em Off center: The Republican revolution and the erosion of
  American democracy}.
\newblock Yale University Press, 2006.

\bibitem{theriault2013gingrich}
Sean~M Theriault.
\newblock {\em The Gingrich senators: The roots of partisan warfare in
  Congress}.
\newblock Oxford University Press, 2013.

\bibitem{faris2017partisanship}
Robert Faris, Hal Roberts, Bruce Etling, Nikki Bourassa, Ethan Zuckerman, and
  Yochai Benkler.
\newblock Partisanship, propaganda, and disinformation: Online media and the
  2016 us presidential election.
\newblock {\em Berkman Klein Center Research Publication}, 6, 2017.

\bibitem{brady2019ideological}
William~J Brady, Julian~A Wills, Dominic Burkart, John~T Jost, and Jay~J
  Van~Bavel.
\newblock An ideological asymmetry in the diffusion of moralized content on
  social media among political leaders.
\newblock {\em Journal of Experimental Psychology: General}, 148(10):1802,
  2019.

\bibitem{gant}
Michael~M Gant and Dwight~F Davis.
\newblock {Mental Economy and Voter Rationality: The Informed Citizen Problem
  in Voting Research}.
\newblock {\em The Journal of Politics}, 46(1):132--153, 1984.

\bibitem{Stucki2018}
Iris Stucki, Lyn~E. Pleger, and Fritz Sager.
\newblock {The Making of the Informed Voter: A Split-Ballot Survey on the Use
  of Scientific Evidence in Direct-Democratic Campaigns}.
\newblock {\em Swiss Political Science Review}, 24(2):115--139, jun 2018.

\bibitem{ashworth}
Scott Ashworth and Ethan Bueno D~E Mesquita.
\newblock {Is Voter Competence Good for Voters?: Information, Rationality, and
  Democratic Performance}.
\newblock {\em The American Political Science Review}, 108(3):565--587, 2014.

\bibitem{aiello2012people}
Luca~Maria Aiello, Martina Deplano, Rossano Schifanella, and Giancarlo Ruffo.
\newblock People are strange when you’re a stranger: Impact and influence of
  bots on social networks.
\newblock {\em Proceedings of the International AAAI Conference on Web and
  Social Media}, 6(1), May 2012.

\bibitem{zelenkauskaite2017information}
Asta Zelenkauskaite and Marcello Balduccini.
\newblock “information warfare” and online news commenting: Analyzing
  forces of social influence through location-based commenting user typology.
\newblock {\em Social Media + Society}, 3(3):2056305117718468, 2017.

\bibitem{yang2019arming}
Kai-Cheng Yang, Onur Varol, Clayton~A. Davis, Emilio Ferrara, Alessandro
  Flammini, and Filippo Menczer.
\newblock Arming the public with artificial intelligence to counter social
  bots.
\newblock {\em Human Behavior and Emerging Technologies}, 1(1):48--61, 2019.

\bibitem{sayyadiharikandeh2020detection}
Mohsen Sayyadiharikandeh, Onur Varol, Kai-Cheng Yang, Alessandro Flammini, and
  Filippo Menczer.
\newblock Detection of novel social bots by ensembles of specialized
  classifiers.
\newblock {\em Proceedings of the 29th ACM International Conference on
  Information \& Knowledge Management}, Oct 2020.

\bibitem{yang2020scalable}
Kai-Cheng Yang, Onur Varol, Pik-Mai Hui, and Filippo Menczer.
\newblock Scalable and generalizable social bot detection through data
  selection.
\newblock {\em Proceedings of the AAAI Conference on Artificial Intelligence},
  34(01):1096--1103, Apr. 2020.

\bibitem{davis2016osome}
Clayton~A Davis, Giovanni~Luca Ciampaglia, Luca~Maria Aiello, Keychul Chung,
  Michael~D Conover, Emilio Ferrara, Alessandro Flammini, Geoffrey~C Fox,
  Xiaoming Gao, Bruno Gon{\c{c}}alves, et~al.
\newblock Osome: the iuni observatory on social media.
\newblock {\em PeerJ Computer Science}, 2:e87, 2016.

\bibitem{ferrara2019history}
Emilio Ferrara.
\newblock The history of digital spam.
\newblock {\em Communications of the ACM}, 62(8):82--91, 2019.

\bibitem{Ferrara2020}
Emilio Ferrara, Herbert Chang, Emily Chen, Goran Muric, and Jaimin Patel.
\newblock {Characterizing social media manipulation in the 2020 U.S.
  presidential election}.
\newblock {\em First Monday}, oct 2020.

\bibitem{Ferrara2020_1}
Emilio Ferrara.
\newblock What types of covid-19 conspiracies are populated by twitter bots?
\newblock {\em First Monday}, 2020.

\bibitem{jiang2020political}
Julie Jiang, Emily Chen, Shen Yan, Kristina Lerman, and Emilio Ferrara.
\newblock Political polarization drives online conversations about covid-19 in
  the united states.
\newblock {\em Human Behavior and Emerging Technologies}, 2(3):200--211, 2020.

\bibitem{bail2018exposure}
Christopher~A. Bail, Lisa~P. Argyle, Taylor~W. Brown, John~P. Bumpus, Haohan
  Chen, M.~B.~Fallin Hunzaker, Jaemin Lee, Marcus Mann, Friedolin Merhout, and
  Alexander Volfovsky.
\newblock Exposure to opposing views on social media can increase political
  polarization.
\newblock {\em Proceedings of the National Academy of Sciences},
  115(37):9216--9221, 2018.

\bibitem{Allcott2017}
Hunt Allcott and Matthew Gentzkow.
\newblock {Social media and fake news in the 2016 election}, mar 2017.

\bibitem{VanProoijen2019}
Jan-Willem van Prooijen.
\newblock {Belief in Conspiracy Theories}.
\newblock In {\em The Social Psychology of Gullibility}, pages 319--332.
  Routledge, apr 2019.

\bibitem{Douglas2019}
Karen~M. Douglas, Joseph~E. Uscinski, Robbie~M. Sutton, Aleksandra Cichocka,
  Turkay Nefes, Chee~Siang Ang, and Farzin Deravi.
\newblock {Understanding Conspiracy Theories}.
\newblock {\em Political Psychology}, 40(S1):3--35, feb 2019.

\bibitem{Byford2011}
Jovan Byford.
\newblock {\em {Conspiracy Theories - A critical Intoduction}}.
\newblock 2011.

\bibitem{Zuckerman2019}
Ethan Zuckerman.
\newblock {QAnon and the Emergence of the Unreal}.
\newblock {\em Journal of Design and Science}, (6), jul 2019.

\bibitem{Stella2018}
Massimo Stella, Emilio Ferrara, and Manlio {De Domenico}.
\newblock {Bots increase exposure to negative and inflammatory content in
  online social systems}.
\newblock {\em Proceedings of the National Academy of Sciences of the United
  States of America}, 115(49):12435--12440, dec 2018.

\bibitem{rosenblum2020lot}
Nancy~L Rosenblum and Russell Muirhead.
\newblock {\em A lot of people are saying: The new conspiracism and the assault
  on democracy}.
\newblock Princeton University Press, 2020.

\bibitem{paul2016russian}
Christopher Paul and Miriam Matthews.
\newblock The russian “firehose of falsehood” propaganda model.
\newblock {\em Rand Corporation}, pages 2--7, 2016.

\bibitem{lewandowsky2012misinformation}
Stephan Lewandowsky, Ullrich~KH Ecker, Colleen~M Seifert, Norbert Schwarz, and
  John Cook.
\newblock Misinformation and its correction: Continued influence and successful
  debiasing.
\newblock {\em Psychological science in the public interest}, 13(3):106--131,
  2012.

\bibitem{vosoughi2018spread}
Soroush Vosoughi, Deb Roy, and Sinan Aral.
\newblock The spread of true and false news online.
\newblock {\em Science}, 359(6380):1146--1151, 2018.

\bibitem{tangherlini2020automated}
Timothy~R Tangherlini, Shadi Shahsavari, Behnam Shahbazi, Ehsan Ebrahimzadeh,
  and Vwani Roychowdhury.
\newblock An automated pipeline for the discovery of conspiracy and conspiracy
  theory narrative frameworks: Bridgegate, pizzagate and storytelling on the
  web.
\newblock {\em PloS one}, 15(6):e0233879, 2020.

\bibitem{muis2017causes}
Jasper Muis and Tim Immerzeel.
\newblock Causes and consequences of the rise of populist radical right parties
  and movements in europe.
\newblock {\em Current Sociology}, 65(6):909--930, 2017.

\bibitem{zhuravskaya2020political}
Ekaterina Zhuravskaya, Maria Petrova, and Ruben Enikolopov.
\newblock Political effects of the internet and social media.
\newblock {\em Annual Review of Economics}, 12:415--438, 2020.

\end{thebibliography}

\end{document}